\documentclass{article}
\usepackage{amsmath}
\usepackage{amsfonts}
\usepackage{amssymb}
\usepackage[bottom=3.5cm, top=4cm, inner=3cm, outer=3cm]{geometry}
\usepackage{natbib}
\usepackage{multirow}
\usepackage[toc,page]{appendix}

\begin{document}

\begin{center}
{\Large {\bf Empirical and Simulated Adjustments of Composite \\[1ex] Likelihood Ratio Statistics}}
\end{center}

\vspace{1cm}

\noindent Manuela Cattelan\footnote{manuela.cattelan@unipd.it, \\
\hspace*{0.42cm} Department of Statistical Sciences, University of Padova, \\ \hspace*{0.42cm} via C. Battisti 241, 35121 Padova, Italy} \\[1ex]
{\em Department of Statistical Sciences, University of Padua, Italy}

\

\noindent Nicola Sartori\\[1ex]
{\em Department of Statistical Sciences, University of Padua, Italy}

\vspace{1cm}

\noindent
{\bf Abstract} Composite likelihood inference has gained much popularity thanks to its computational manageability and its theoretical properties. Unfortunately, performing composite likelihood ratio tests is inconvenient because of their awkward asymptotic distribution. There are many proposals for adjusting composite likelihood ratio tests in order to recover an asymptotic chi square distribution, but they all depend on the sensitivity and variability matrices. The same is true for Wald-type and score-type counterparts. In realistic applications sensitivity and variability matrices usually need to be estimated, but there are no comparisons of the performance of composite likelihood based statistics in such an instance. A comparison of the accuracy of inference based on the statistics considering two methods typically employed for estimation of sensitivity and variability matrices, namely an empirical method that exploits independent observations, and Monte Carlo simulation, is performed. The results in two examples involving the pairwise likelihood show that a very large number of independent observations should be available in order to obtain accurate coverages using empirical estimation, while limited simulation from the full model provides accurate results regardless of the availability of independent observations.

\


\section{Introduction}
The use of the likelihood function to perform inference in statistical models is becoming more and more cumbersome for diverse reasons, as for example the availability of huge datasets and the implementation of complex models developed to reproduce natural phenomena. This problem is often overcome through the definition of pseudo-likelihood functions that are computationally manageable, but retain some nice properties of the likelihood function. Many of the pseudo-likelihood functions proposed in the literature belong to the class of composite likelihoods \citep{Lindsay:88, Varin:08, Varin:11}. Indeed, the definition of composite likelihood given by \cite{Lindsay:88} is quite general and encompasses any function which is a product of marginal or conditional probabilities for subsets of events. Composite likelihoods share some nice properties of the ordinary likelihood, as the unbiasedness of the composite likelihood score function and the asymptotic normal distribution of the maximum composite likelihood estimator \citep{Molenberghs:05}. The simplifications of both computational issues and model assumptions that derive from this type of pseudo-likelihood led to a considerable diffusion of composite likelihood estimation and the consequent  investigation of its theoretical properties and the development of further inferential techniques based on composite likelihood. 

In this paper, we focus on hypothesis testing and confidence regions construction when a composite likelihood is employed. 
There are composite likelihood versions of the tests developed in the full likelihood context. Hence, Wald-type, score-type and likelihood ratio statistics based on the composite likelihood can be specified. However, as with the full likelihood, the Wald-type statistic lacks invariance under reparameterisations of the model and forces confidence regions to have an elliptical shape. On the other hand, score-type statistics are often numerically unstable \citep{Rotnitzky:90, Molenberghs:05, Pace:11}, while composite likelihood ratio statistics do not have the usual asymptotic chi square distribution. 

There are different proposals to overcome the problem of the awkward asymptotic distribution of the composite likelihood ratio statistic. All such proposals, as well as the Wald-type and score-type statistics, depend on sensitivity and variability matrices, which are, respectively, the expected value of minus the hessian of the composite log likelihood and the variance of the composite score function. The computation of these matrices is generally cumbersome and approximations are typically used \citep[\S 5.1]{Varin:11}. The main purpose of this paper is to compare the behavior of the various statistics when they are based on estimated sensitivity and variability matrices. In particular, empirical and Monte Carlo estimates are considered. Two simulation studies are implemented in order to compare the performance of adjusted composite likelihood ratio statistics when pairwise likelihood is used for inferential purposes.

The paper is organized as follows. Section \ref{sec2} reviews composite likelihood based statistics and the proposals to overcome the problem of the asymptotic distribution of the composite likelihood ratio statistics. Section \ref{sec3} presents the methods commonly employed to estimate the sensitivity and variability matrices. Section \ref{sec4} shows the results of simulation studies that compare the different statistics in two model settings, namely a spatial Gaussian random field and a multivariate probit model, and Section \ref{sec5} concludes with a discussion.

\section{Adjusting composite likelihood ratio statistics}
\label{sec2}
Let $y_{1}, \ldots, y_{n}$ be independent realizations of a $q$-dimensional random vector $Y_{i}=(Y_{i1}, \ldots, Y_{iq})$, with density or probability function $f(y_{i}; \theta)$ depending on a $d$-dimensional parameter $\theta$. If the full likelihood is computationally cumbersome, or the model cannot be fully specified, a composite likelihood may offer a valid alternative. A composite likelihood is a combination of likelihoods for conditional or marginal events \citep{Lindsay:88}. Assume there are $K$ marginal or conditional events $A_{k}(y_{i})$ involving elements of $y_{i}$, $k=1, \ldots, K$, for which we can compute the likelihood $L_{k}(\theta; y_{i}) \propto f(Y_{i} \in A_{k}; \theta)$, then the composite likelihood is
$$
cL(\theta; y)= \prod_{i=1}^{n} \prod_{k=1}^{K} L_{k}(\theta; y_{i})^{w_{k}},
$$
where $w_{k}$ are non negative weights and $y=(y_{1}, \ldots, y_{n})$. The composite log likelihood is $cl(\theta; y)= \log cL(\theta; y)$ and the composite score function is $cU(\theta; y)=\nabla_{\theta} \, cl(\theta; y)$. The maximizer of $cl(\theta;y)$, $\hat \theta_{c}$, is the maximum composite likelihood estimate. 

Under fairly general regularity conditions the maximum composite likelihood estimator is asymptotically normally distributed, $\hat \theta_{c} \stackrel{\cdot}{\sim} N_{d}(\theta, G(\theta)^{-1})$, where $G(\theta)$ denotes the Godambe information matrix. Specifically, the asymptotic covariance matrix is $G(\theta)^{-1}=H(\theta)^{-1} J(\theta) H(\theta)^{-1}$, where $H(\theta)= \text{E}\{-\nabla_{\theta} cU(\theta;y)\}$ is called the sensitivity matrix and $J(\theta)= \text{E}\{cU(\theta) cU(\theta)^{T}\}$ is called the variability matrix. The composite likelihood is not a proper likelihood, but it can be interpreted as the likelihood for a misspecified model; as a consequence the second Bartlett identity does not hold and typically $J(\theta) \neq H(\theta)$. 

A type of composite likelihood often used in applications is the pairwise likelihood, which is the product of marginal bivariate probabilities,
$$
pL(\theta; y)= \prod_{i=1}^n \prod_{j=1}^{q-1} \prod_{k=j+1}^q f(y_{ij}, y_{ik}; \theta)^{w_{ij,ik}},
$$
and the pairwise log likelihood is $pl(\theta; y)=\log pL(\theta; y)$.

Assume that interest lies in a $p$-dimensional parameter $\gamma$, where $\theta=(\gamma, \delta)$ and $\delta$ is a nuisance parameter of dimension $d-p$. It is possible to define test statistics based on the composite likelihood which are analogous to those based on the full likelihood. Denote by $\hat \theta_{c \gamma}$ the constrained maximum composite likelihood estimate of $\theta$ for a fixed $\gamma$, and let $\hat \theta_c=(\hat \gamma_c, \hat \delta_c)$. The Wald-type statistic for the parameter of interest is 
\begin{equation}
\label{cW}
cW(\gamma)=(\hat \gamma_{c} - \gamma)^{T} \{G^{\gamma \gamma}(\hat \theta_{c \gamma})\}^{-1} (\hat \gamma_{c} - \gamma),
\end{equation}
where $G^{\gamma \gamma}(\hat \theta_{c \gamma})$ is the $p \times p$ submatrix of the inverse of $G(\hat \theta_{c \gamma})$ pertaining to $\gamma$. The statistic $cW(\gamma)$ has an asymptotic $\chi^{2}_{p}$ distribution. Unfortunately, this quantity is not invariant to reparameterisations of the model.

The score-type statistic based on the composite likelihood is
\begin{equation}
\label{cS}
cS(\gamma)=cU_{\gamma}(\hat \theta_{c \gamma}) H^{\gamma \gamma}(\hat \theta_{c \gamma}) \{G^{\gamma \gamma}(\hat \theta_{c \gamma}) \}^{-1} H^{\gamma \gamma} (\hat \theta_{c \gamma}) cU_{\gamma}(\hat \theta_{c \gamma}),
\end{equation}
where $cU_{\gamma}(\theta)= \nabla_\gamma cl(\theta;y)$ is the derivative of the composite log likelihood with respect to the parameter of interest and $H^{\gamma \gamma}(\hat \theta_{c \gamma})$ denotes the submatrix of the inverse of $H(\hat \theta_{c \gamma})$ pertaining to $\gamma$. The asymptotic distribution of $cS(\gamma)$ is $\chi^2_p$, but this statistic is often numerically unstable \citep{Molenberghs:05}.

Finally, it is possible to define also a composite likelihood ratio statistic
$$
cLR(\gamma)=2 \{cl(\hat \theta_{c})- cl(\hat \theta_{c \gamma}) \}.
$$
Its asymptotic distribution is a weighted sum of $p$ independent chi square random variables with one degree of freedom, precisely $\sum_{i=1}^{p} \omega_{i} \chi^{2}_{1i}$, where $\omega_{1}, \ldots, \omega_{p}$ are the eigenvalues of $\{H^{\gamma \gamma}(\theta)\}^{-1} G^{\gamma \gamma}( \theta)$. These can be consistently estimated by evaluating the matrices in $\hat \theta_{c \gamma}$.
This awkward distribution prevents the use of the composite likelihood ratio statistic when the dimension of the parameter of interest is larger than one. For this reason various adjustments have been proposed, mainly in order to recover an approximate $\chi^2_p$ distribution.

A first proposal for the adjustment of composite likelihood ratio statistics suggests to match the first order moment of the composite likelihood ratio statistic with that of a $\chi^{2}_{p}$ random variable \citep{Molenberghs:05}
$$
cLR(\gamma)_{1}= \overline \omega^{\, -1} cLR(\gamma),
$$
where $\overline \omega= \sum_{i=1}^{p} \omega_{i}/p$, and then use a $\chi^{2}_{p}$ as approximate distribution. A better approximation can be obtained through first and second order moment matching \citep{Varin:08}, which gives a Satterthwaite type adjustment \citep{Satterthwaite:46}
$$
cLR(\gamma)_{2}= \kappa^{-1} cLR(\gamma),
$$
where $\kappa= \sum_{i=1}^{p} \omega_{i}^{2}/ \sum_{i=1}^{p} \omega_{i}$. This quantity has an asymptotic $\chi^{2}_{\nu}$ distribution, where the degrees of freedom are $\nu=(\sum_{i=1}^{p} \omega_{i})^{2}/\sum_{i=1}^{p} \omega_{i}^{2}$. Quantities $\overline \omega$, $\kappa$ and $\nu$ depend on $\theta$ and are usually evaluated at $\hat \theta_{c \gamma}$. The improved accuracy of $cLR(\gamma)_{2}$ on $cLR(\gamma)_{1}$ is counterbalanced by the inconvenient dependence of its asymptotic distribution on the parameter $\gamma$. 

Other two adjustments of the composite likelihood ratio statistics are proposed by \cite{Chandler:07} and \cite{Pace:11}. The former authors suggest the following adjusted statistic
\begin{equation}
\label{cCB}
cLR(\gamma)_{CB}=\frac{(\hat\gamma_{c}-\gamma)^{T} \{G^{\gamma \gamma}(\hat \theta_c)\}^{-1} (\hat\gamma_{c}-\gamma)}{(\hat\gamma_{c}-\gamma)^{T}H_{\gamma \gamma}(\hat \theta_c) (\hat\gamma_{c}-\gamma)^{T}} cLR(\gamma),
\end{equation}
which has asymptotic $\chi^{2}_{p}$ distribution. 
In a simulation study \cite{Chandler:07} show that their proposal behaves well, and at least it does not perform worse than statistics (\ref{cW}) and (\ref{cS}) in all settings considered. However, \cite{Pace:11} show that $cLR(\gamma)_{CB}$ is not parameterisation invariant, and therefore propose a different rescaling that preserves the parameterisation invariance of the likelihood ratio statistic, that is
 \begin{equation}
 \label{cINV}
 cLR(\gamma)_{I}=\frac{cS(\gamma)}{cU_{\gamma}(\hat \theta_{c \gamma}) H^{\gamma \gamma}(\hat \theta_{c \gamma}) cU_{\gamma}(\hat \theta_{c \gamma}) } cLR(\gamma),
 \end{equation}
which is again asymptotically $\chi^{2}_{p}$ distributed. Despite being partially based on the score statistic $cS(\gamma)$, $cLR(\gamma)_{I}$ usually does not inherit its numerical instability.

The performance of the different adjustments is compared in a simulation study in \cite{Pace:11} that consider two different model settings: equicorrelated multivariate normal data and first order autoregression. In both cases, the authors use pairwise likelihood for making inference on model parameters and compare the results with those produced by maximum likelihood based statistics. Moreover, in both settings it is possible to compute analytically the Fisher information matrix and the matrices $H(\theta)$ and $J(\theta)$ for the pairwise likelihood. In general, the statistic (\ref{cINV}) seems to behave well in all settings considered, while in some instances the empirical coverage of adjustment (\ref{cCB}) is much lower than the nominal value.
These results are obtained when the quantities of interest can be computed analytically. This rarely occurs in applications where composite likelihood is employed. Indeed, composite likelihood is often used in complex models where not only it is not possible to deal with the full likelihood, but also the analytical computation of $H(\theta)$ and $J(\theta)$ is typically unfeasible. The main concern of this paper is to investigate the behavior of the different proposals when the quantities involved in the computation of the statistics have to be estimated.

\section{Estimation of $H(\theta)$ and $J(\theta)$}
\label{sec3}

Estimation of the matrices $H(\theta)$ and $J(\theta)$ is a typical concern in applications in which composite likelihood is employed since they are necessary ingredients also for the computation of the standard errors of the maximum composite likelihood estimates. While $H(\theta)$ can be reasonably estimated through the observed hessian, the estimation of the variability matrix $J(\theta)$ poses major difficulties.

Matrices $H(\theta)$ and $J(\theta)$ are usually estimated either empirically, exploiting groups of independent or almost independent data, or through simulation. 
When there are groups of independent observations, as for example when data are divided into clusters, it is possible to estimate $J(\theta)$ as 
$$
\hat J^{E}(\theta)= \frac{1}{n} \sum_{i=1}^{n} cU( \theta; y_{i}) cU( \theta; y_{i})^{T},
$$
where $cU( \theta; y_{i}) $ denotes the elements of the composite score involving only observations of the vector $y_{i}$. 
For example, $cU( \theta; y_{i})= \sum_{j=1}^{q-1} \sum_{k=j+1}^{q} cU(\theta; y_{ij}, y_{ik})=  \sum_{j=1}^{q-1} \sum_{k=j+1}^{q} w_{ij,ik}  \nabla _{\theta} \log f(y_{ij}, y_{ik}; \theta)$ if pairwise likelihood is employed. When independent repetitions of the data are not available, as often happens in time series or spatial data, but it is possible to identify groups of data with low dependence, this method may be applied to groups of slightly dependent data. For example, when dealing with time series with dependence decreasing in time, a window subsampling method may be employed \citep{Varin:08}.

The empirical estimate of the sensitivity matrix is
$$
\hat H^{E}(\theta)=- \frac{1}{n} \sum_{i=1}^{n} \nabla_{\theta} cU( \theta; y_{i}),
$$
which corresponds to minus the Hessian matrix. However, since the second Bartlett identity holds for single subsets of the data \citep{Varin:08}, the sensitivity matrix can also be estimated as
$$
\hat H^{E}(\theta)= \frac{1}{n} \sum_{i=1}^{n} \sum_{k=1}^{K} cU( \theta; y_{i} \in A_{k}) cU(\theta; y_{i} \in A_{k}) ^{T},
$$
which avoids the computation of the second derivative.
When pairwise likelihood is employed, this corresponds to 
$$
\hat H^{E}(\theta)= \frac{1}{n} \sum_{i=1}^{n} \sum_{j=1}^{q-1} \sum_{k=j+1}^{q} cU(\theta; y_{ij}, y_{ik}) cU( \theta; y_{ij}, y_{ik})^{T}.
$$
The empirical estimation of $H(\theta)$ and $J(\theta)$ does not require any further assumption than those made for the composite likelihood function, which consist only in the specification of low order marginal or conditional probabilities.

An alternative method to estimate the Godambe information matrix is through simulation, which requires assumptions about the full distribution of the data. Such assumptions are not always possible; for example \cite{Xu:11} consider a model that has multivariate normal marginals but which is not jointly multivariate normally distributed, or \cite{Cattelan:13} introduce a Bradley-Terry-Dale model for which the specification of the multivariate distribution is theoretically possible, but in practice extremely difficult, hence only marginal bivariate distributions can be considered. However, although the assumption about the full distribution of the data may appear an important limitation of this method, in most of the applications of composite likelihood a full model is assumed for the data, but the difficulties in computing the likelihood function lead to the use of a composite likelihood. In these cases the likelihood function is difficult to evaluate, but it may be straightforward to simulate from the full model, as happens in modern Approximate Bayesian Computation methods, which are nowadays widely used \citep{Marin:12}. 

Let $y^{m}$, $m=1, \ldots, M$, denote the $m$th dataset simulated from $f(y; \theta)$, the full distribution of the data. Then, the Monte Carlo estimates of $J(\theta)$ and $H(\theta)$ are
$$
\hat J^{S}(\theta)= \frac{1}{M} \sum_{m=1}^{M}  cU(\theta; y^{m}) cU(\theta; y^{m})^{T},
$$
and
$$
\hat H^{S}(\theta) = -\frac{1}{M} \sum_{m=1}^{M}  \nabla_{\theta} cU( \theta; y^{m}).
$$
Again, in the estimation of $H(\theta)$ it is possible to exploit the second Bartlett identity, which is valid for each component of the composite likelihood. This may be convenient especially if analytical first derivatives are available. As will be shown in the next section, a few hundred simulated datasets are typically sufficient for reasonable accuracy.

Even when it is possible to compute $J(\theta)$ exactly, it may be computationally more convenient to use $\hat J^{S}(\theta)$. Indeed, consider a single observation ($n=1$) of a $q$-dimensional multivariate normal random vector, as in many applications in spatial statistics. The computational cost of the likelihood is of order $O(q^{2.81})$, while that of the pairwise likelihood and score functions is of order $O(q^{2})$. On the other hand, the computational cost of $J(\theta)$ is of order $O(q^{4})$, while that of $\hat J^{S}(\theta)$ is $O(M q^{2})$; for an example see Section \ref{sec4.1}.

The main interest here lies in investigating whether there are differences in the performances of the various composite likelihood based statistics when $H(\theta)$ and $J(\theta)$ have to be estimated with respect to cases in which they are available analytically, and which of the two estimating methods yields better results. Such an investigation has an important practical relevance since the estimation of $H(\theta)$ and $J(\theta)$ is the only option in most realistic applications. The proposed solutions are explored in simulation studies.

\section{Simulation studies}
\label{sec4}
Simulation studies are performed considering two different models and using pairwise likelihood for inferential purposes. The first model assumes a Gaussian random field, which is often employed in spatial statistics. In this case it is possible to compute analytically the sensitivity and the variability matrices, thus allowing a comparison of the performance of analytical, empirical and simulation based quantities. Instead, the second model considered is a multivariate probit model, in which the analytical form of $H(\theta)$ and $J(\theta)$ is not available. 

\subsection{Spatial Gaussian random field}
\label{sec4.1}
Let $Y_{i}$, $i=1, \ldots, n$, be independent random vectors following a $q$-di\-men\-sion\-al normal distribution with mean $\mu 1_{q}$, with $1_{q}$ a vector of ones of length $q$, and a stable covariance matrix, $\text{cov}(Y_{ij}, Y_{i k})=\sigma^{2} \exp\left\{-\left(\frac{d_{j k}}{\lambda}\right)^{\alpha}\right\}$, where $d_{j k}$ denotes the distance between the spatial locations in which $Y_{ij}$ and $Y_{ik}$ are measured, $\lambda>0$ and $\alpha \in (0,2]$. Couples of observations $(Y_{ij}, Y_{ik})$ have a bivariate normal distribution with components with mean $\mu$, variance $\sigma^{2}$ and correlation $\rho_{jk}= \exp\left\{-\left(\frac{d_{j k}}{\lambda}\right)^{\alpha}\right\}$. Thus the pairwise log likelihood is
\begin{eqnarray*}
pl (\theta; y)&=&\sum_{i=1}^n\sum_{j=1}^{q-1} \sum_{k=j+1}^{q} w_{ij,ik}\log f(Y_{ij}=y_{ij},Y_{ik}=y_{ik};\theta)\\
& = & \sum_{i=1}^n \sum_{j=1}^{q-1}\sum_{k=j+1}^q w_{ij,ik}\left[
-\log\sigma^2 -\frac{1}{2}\log(1-\rho^2_{jk}) -\frac{A_{ijk}}{2\sigma^2(1-\rho^2_{jk}) }\right],
\end{eqnarray*}
where 
$
A_{ijk}=(y_{ij}-\mu)^2+ (y_{ik}-\mu)^2-2\rho_{jk}(y_{ij}-\mu)(y_{ik}-\mu)\,.
$

We assume that the independent replications $Y_{i}$, $i=1, \ldots, n$, are in the same spatial locations. Therefore, since the weights typically depend on the distance $d_{jk}$, we have $w_{ij, ik}=w_{jk}$. Typical choices of the weights are decreasing functions of the distance, or dichotomous weights such that $w_{jk}=1$ if $d_{jk}$ is lower than a given threshold $d_{0}$, and $w_{jk}= 0$ otherwise. The choice of the weights might affect the efficiency of the pairwise likelihood estimates. This aspect is still an open problem and its investigation is outside the scope of the paper; see \cite{Bevilacqua:14} and references therein. 

The Gaussian random field model is particularly appealing since it is possible to compute the quantities $H(\theta)$ and $J(\theta)$ analytically. Hence, the coverages of the tests based on analytical quantities can be compared to those of the tests based on empirical or simulated matrices. The analytical forms of $H(\theta)$ and $J(\theta)$ are given in Appendix A. Moreover, it is possible to perform ordinary maximum likelihood estimation of this model, therefore also the performance of the likelihood ratio test is available. 

The components of the pairwise score function are 
\begin{eqnarray*}
\frac{\partial pl(\theta; y)}{\partial \mu} &=&\sum_{i=1}^{n} \sum_{j=1}^{q-1} \sum_{k=j+1}^{q} w_{jk} \frac{1}{ \sigma^{2}(1+ \rho_{jk})} [(y_{ij}-\mu)+(y_{ik}-\mu)],\\
\frac{\partial pl(\theta;y)}{\partial \sigma^{2}} &=&\sum_{i=1}^{n} \sum_{j=1}^{q-1} \sum_{k=j+1}^{q} w_{jk} \left[  -\frac{1}{\sigma^{2}} + \frac{A_{ijk}}{2 (\sigma^{2})^{2}(1-\rho_{jk}^{2})} \right],\\
\frac{\partial pl(\theta; y)}{\partial \gamma} &=&\sum_{i=1}^{n} \sum_{j=1}^{q-1} \sum_{k=j+1}^{q} w_{jk} \frac{\partial \rho_{jk}}{\partial \gamma} \frac{1}{1-\rho_{jk}^{2}}\left[  \rho_{jk} -\frac{\rho_{jk} A_{ijk}}{\sigma^{2}(1-\rho_{jk}^{2})} + \frac{(y_{ij}-\mu)(y_{ik}-\mu)}{\sigma^{2}}\right],\\
\end{eqnarray*}
where $\gamma=(\lambda, \alpha)$ and $\partial \rho_{jk}/ \partial \gamma =(\partial \rho_{jk}/ \partial \lambda, \partial \rho_{jk}/ \partial \alpha)^{T}$, specifically 

\begin{eqnarray*}
\frac{\partial \rho_{jk}}{\partial \lambda}&=& \alpha \, \frac{\rho_{jk}}{\lambda} \left(\frac{d_{jk}}{\lambda}\right)^{\alpha}, \\
\frac{\partial \rho_{jk}}{ \partial \alpha}&=& \rho_{jk}\left(\frac{d_{jk}}{\lambda}\right)^{\alpha} \left[-\log\left(\frac{d_{jk}}{\lambda}\right)\right].
\end{eqnarray*}

\begin{table}[bt]
\begin{center}
{\footnotesize
\begin{tabular}{cccccccc}
$n$		& $LRT$ 	&	$LR_{2}^{A}$ & 	$LR_{2}^{E}$	& $LR_{2}^{S}$	& $LR_{I}^{A}$	& $LR_{I}^{E}$	&	$LR_{I}^{S}$\\
\hline
\multicolumn{8}{c}{95.0}\\
\hline
1 & 97.1  & 99.7 &  -& 99.8 & 98.4 &  -& 98.4 \\ 
  5 & 95.1 &  97.3 & 84.5 & 97.5 & 96.9 & 87.5 & 96.9 \\ 
  30 & 94.8 &  95.3 & 89.1 & 95.5 & 95.1 & 84.3 & 95.2 \\ 
  	\hline			
\multicolumn{8}{c}{99.0}\\
\hline
  1 & 99.7  & 99.9 &  -& 100.0 & 99.9 & - & 99.9 \\ 
  5 & 99.1 &  99.4 & 98.7 & 99.3 & 99.3 & 94.4 & 99.3 \\ 
  30 & 98.9  & 99.0 & 96.0 & 99.1 & 99.1 & 92.3 & 99.1 \\ 
   \hline
\end{tabular}}
\caption{Empirical coverages of the statistics: likelihood ratio test based on the ordinary log likelihood ($LRT$), composite likelihood ratio using second order matching adjustment ($LR_{2}$) and composite likelihood ratio adjustment by \cite{Pace:11} ($LR_{I}$) for nominal values 95\% and 99\% in a spatial Gaussian random field for parameter of interest $(\lambda, \alpha)$, with $n=1, 5, 30$, using analytical ($^{A}$), empirical ($^{E}$) and Monte Carlo ($^{S}$) versions of $H(\theta)$ and $J(\theta)$.}
\label{tab1}
\end{center}
\end{table}

\begin{table}[tb]
\begin{center}
\footnotesize{
\begin{tabular}{rrrrrrrr}
  \hline
  	&	 \multicolumn{3}{c}{$LR_{2}^{S}$} && \multicolumn{3}{c}{$LR_{I}^{S}$}\\
	\cline{2-4} \cline{6-8} 
 n & 1 & 5 & 30 && 1 & 5 & 30 \\ 
  \hline
  M	& \multicolumn{7}{c}{95.0}\\
  \hline
100 &  99.6 & 97.4 & 95.5 && 98.0 & 96.4 & 95.0 \\ 
  250 &  99.7 & 97.6 & 95.4 && 98.3 & 96.7 & 95.0 \\ 
  500 &  99.7 & 97.6 & 95.5 && 98.3 & 96.9 & 95.1 \\ 
    \hline
  	& \multicolumn{7}{c}{99.0}\\
  \hline
  100 &  100.0 & 99.3 & 99.1 && 99.8 & 99.2 & 98.9 \\ 
  250 &  99.9 & 99.5 & 99.1 && 99.8 & 99.3 & 98.9 \\ 
  500 &  100.0 & 99.4 & 99.2 && 99.8 & 99.3 & 99.1 \\ 
   \hline
\end{tabular}}
\caption{Comparison of coverages of the statistics: composite likelihood ratio using second order matching adjustment ($LR_{2}$) and composite likelihood ratio adjustment by \cite{Pace:11} ($LR_{I}$) based on Monte Carlo simulation as $M$ increases in a Gaussian random field.}
\label{tabM1}
\end{center}
\end{table}

A simulation study is performed considering the correlation parameters $\lambda$ and $\alpha$ as parameters of interest and $\mu$ and $\sigma^{2}$ as nuisance parameters. In each setting 10,000 data sets are simulated on a regular square grid between 0 and 7, so each observation has dimension $q=64$. The values for the parameters are $\mu=0$, $\sigma^{2}=2$, $\lambda=0.7$ and $\alpha=1$. In the pairwise likelihood $d_{0}=3$ is used and $n=1, 5$ and $30$ independent replications are considered. Typically, in spatial applications no independent repetitions of the data are available, thus $n=1$. In such cases, the empirical computation of $H(\theta)$ and $J(\theta)$ is not possible. However, in some applications, subgroups of the data are considered as independent and they are employed to compute empirical versions of $H(\theta)$ and $J(\theta)$. This method may be expected to yield less accurate results since the data are not independent. Moreover, it requires a criterion to choose the dimension of the subgroups, and this is not straightforward. 

In order to obtain an accurate approximation of the sensitivity and the variability matrices, $M=1,000$ simulations are employed for the Monte Carlo estimation of $H(\theta)$ and $J(\theta)$. However, a few hundred repetitions are usually enough, as will be shown in a further simulation study reported later.
Table \ref{tab1} shows the empirical coverages of the likelihood ratio test based on the full likelihood and of the adjustments based on the second order matching and the proposal by \cite{Pace:11}. The coverages of the Wald-type and score-type statistics are reported in Appendix A, while the adjustment (\ref{cCB}) and that based on first order matching are not reported given their poor performance. The accuracy of the coverages of the statistics based on analytical quantities increases as the number of independent observations increases. Coverages of the statistics based on simulated quantities are very close to those obtained from analytical calculations, while the coverages of the statistics based on empirical quantities appear unsatisfactory, even when $n=30$.

In order to obtain accurate estimates of $H(\theta)$ and $J(\theta)$ in the Monte Carlo procedure we employed $M=1,000$ replications. In some instances, this number of replications may require considerable computational time. We therefore investigate whether it is possible to obtain accurate coverages with fewer replications. Table \ref{tabM1} reports the coverages of the statistics for increasing values of replications $M$ and considering dimension $n=1, 5, 30$. In all cases, the results with $M=500$ are almost identical to those obtained with $M=1,000$, and even $M=250$ seems to provide very accurate results. Other numbers of Monte Carlo simulations between 500 and 1,000 yield the same results obtained with $M=500$. The relatively low value of $M$
sufficient for reasonable accuracy may be explained by the fact that matrices $H(\theta)$ and $J(\theta)$ are expected values and they are only a part of the adjusted statistics.

Even though in this context exact computation of $J(\theta)$ and $H(\theta)$ are possible, as anticipated in Section \ref{sec3} it may be computationally more convenient to use a simulation based approach. The two approaches were both implemented in R and C code and, in the same machine, the simulation based method was 12 times faster than the analytical computation of $J(\theta)$ for data simulated on a regular grid $\{0, \ldots, 19\}^{2}$, hence with $q=400$, and using $M=1,000$ Monte Carlo replications. Specifically, the analytical evaluation of the variability matrix took, on average in 100 repetitions, 592 seconds, while the simulated one took 48 seconds.

\subsection{Multivariate probit}

Consider a multivariate probit model in which $Y_{ij}$ is a binary random variable that can assume values either 0 or 1. We use the latent variable representation
$$
Y_{ij}=1 \Leftrightarrow Z_{ij}>0, \hspace{1cm} i=1, \ldots, n, \; \; j=1, \ldots, q,
$$
with $Z_{ij}= x_{ij}^{T} \beta + U_{i} +\epsilon_{ij}$, where $x_{ij}$ is an $r$-dimensional vector of covariates, $\beta$ is a vector of regression parameters, $U_{i} \stackrel{iid}{\sim} N(0, \sigma^{2})$, are independent random effects and $\epsilon_{ij}$ are independent normally distributed errors with mean $0$. The errors are independent of the random effects and their variance is set to 1 for identification purposes. Hence, the latent variables $Z_{ij}$ and $Z_{kl}$ are independent if $i \neq k$, while $Z_{ij}$ and $Z_{ik}$ have correlation $\rho=\sigma^{2}/(1+\sigma^{2})$, $ \forall \, j \neq k$. The full likelihood is cumbersome since it entails calculation of multiple integrals of a $q$-variate multivariate normal distribution. In this instance pairwise likelihood is a valid alternative \citep{Lecessie:94}, indeed the pairwise log likelihood is
\[
pl(\beta,\rho;y) = \sum_{i=1}^{n}\sum_{j=1}^{q-1}\sum_{k=j+1}^{q} w_{ij,ik} \log f(Y_{ij}=y_{ij},Y_{ik}=y_{ik};\beta,\rho), 
\]
where, for instance,  $f(Y_{ij}=1,Y_{ik}=1;\beta,\rho)=\Phi_{2}(\lambda_{ij},\lambda_{ik};\rho)$ is the standard bivariate normal distribution with correlation $\rho$, computed in $(\lambda_{ij}, \lambda_{ik})$ with $\lambda_{ij}=x_{ij}^{T}\beta \sqrt{1- \rho}$. In this context, usually $w_{ij,ik}=1$, $\forall \, i,j,k$.

\begin{table}[bt]
\begin{center}
{\small
\begin{tabular}{c|cccc}
$n$	&	$LR_{2}^{E}$	& $LR_{2}^{S}$	&  $LR_{I}^{E}$	&	$LR_{I}^{S}$\\
\hline
\multicolumn{5}{c}{95.0}\\
\hline
10	&		97.8	&	95.6	&	96.1	&	95.1\\	
30	&		97.2	&	95.0	&	96.0	&	94.7\\
100	&		95.3	&	94.9	&	95.4	&	95.1\\
\hline
\multicolumn{5}{c}{99.0}\\
\hline
10	&		99.5	&	99.1	&	99.0	&	99.1\\
30	&		99.4	&	98.9	&	99.1	&	98.9\\
100	&		99.2	&	98.9	&	98.9	&	98.9\\	
\hline
\end{tabular}}
\end{center}
\caption{Empirical coverages of the statistics: composite likelihood ratio using second order matching adjustment ($LR_{2}$) and composite likelihood ratio adjustment by \cite{Pace:11} ($LR_{I}$) for nominal values 95\% and 99\% for the parameter of interest $(\beta_{1}, \rho)$ in a multivariate probit model with $q=30$ and $n=10, 30, 100$, using empirical ($^{E}$) and Monte Carlo ($^{S}$) versions of $H(\theta)$ and $J(\theta)$.}
\label{tab:covmp1}
\end{table}%

In this case it is not possible to compute analytically the matrices $H(\theta)$ and $J(\theta)$, so only statistics based on estimated quantities can be compared. In the model, an intercept term and one covariate are included. The covariate is simulated from a uniform distribution in $[-1,1]$, while model parameters $(\beta_{0}, \beta_{1}, \sigma^{2})$ are set to $(0.5, 1,1)$. The length of the multivariate binary observations is set to $q=30$, and increasing dimensions of the dataset are considered, namely $n=10, 30$ and $100$. For each setting 10,000 datasets are simulated and the Monte Carlo estimates of $H(\theta)$ and $J(\theta)$ are based on $M=1,000$ replications.

Table \ref{tab:covmp1} shows the empirical coverages when only two parameters are of interest, namely $(\beta_{1}, \rho)$. As expected, for small $n$ the simulation based statistics have better coverages than the empirical based ones and are always quite accurate. However, results in Appendix B show that for the Wald-type statistic  the difference is still evident even with $n=100$.

Finally, Table \ref{tab:covmp2} reports the empirical coverages when $\rho=\sigma^{2}/(1+\sigma^{2})$ is the only parameter of interest. Again, the coverage of simulation based statistics are more accurate for small values of $n$, but when the number of independent repetitions is large, even statistics based on empirical quantities provide accurate results.
The results for the Wald-type and score-type statistics, as well as the coverages when all parameters are of interest, are reported in Appendix B.

\begin{table}[tb]
\begin{center}
{\small
\begin{tabular}{c|cccc}
$n$	&	$LR_{2}^{E}$	& $LR_{2}^{S}$	&  $LR_{I}^{E}$	&	$LR_{I}^{S}$\\
\hline
\multicolumn{5}{c}{95.0}\\
\hline
10	&		97.0	&	95.0	&	97.0	&	95.0	\\
30	&		96.4	&	95.1	&	96.4	&	95.1\\
100	&		95.1	&	95.1	&	95.1	&	95.1\\
\hline
\multicolumn{5}{c}{99.0}\\
\hline
10	&		99.0	&	99.2	&	99.0	&	99.2\\
30	&		99.2	&	99.1	&	99.2	&	99.1\\
100	&		99.0	&	99.0	&	99.0	&	99.0\\	
\hline
\end{tabular}}
\end{center}
\caption{Empirical coverages of the statistics: composite likelihood ratio using second order matching adjustment ($LR_{2}$) and composite likelihood ratio adjustment by \cite{Pace:11} ($LR_{I}$) for nominal values 95\% and 99\% for the parameter of interest $\rho$ in a multivariate probit model with $q=30$ and $n=10, 30, 100$, using empirical ($^{E}$) and Monte Carlo ($^{S}$) versions of $H(\theta)$ and $J(\theta)$.}
\label{tab:covmp2}
\end{table}%

\section{Discussion}
\label{sec5}
This paper considers hypothesis testing using likelihood based statistics when a composite likelihood is employed for inferential purposes. Hypothesis testing presents some difficulties since Wald-type tests lack invariance to reparameterisations of the model, score-type tests are often numerically unstable, while composite likelihood ratio statistics do not follow the usual asymptotic chi square distribution. Many different adjustments of the composite likelihood ratio statistic have been proposed to overcome the problem of its awkward asymptotic distribution. The proposal by \cite{Pace:11} seems an interesting alternative. However its performance has been considered so far only in examples in which the sensitivity and the variability matrices can be computed analytically. This rarely happens in applications in which composite likelihood is employed and typically those matrices need to be estimated. We considered the performance of the different statistics when $H(\theta)$ and $J(\theta)$ are estimated either empirically, or through Monte Carlo simulation. The score-type statistic, the adjustment of the composite likelihood ratio statistic based on second order moment matching and the adjustment proposed by \cite{Pace:11} seem to perform quite well in all situations considered. However, score-type tests can be numerically unstable, while the adjustment based on the second order moment matching has an asymptotic distribution which depends on the parameters of the model. 

The results show that empirical estimation of the sensitivity and variability matrices requires a large number of independent observations and in our simulations it is not very accurate even with a dataset with as much as 100 independent replications. In many applications, as in time series or in spatial statistics, subsets of independent data are not available and the empirical method is applied to subsets of data with low dependence, using for example window subsampling. In these instances we may expect that the performance of the statistics based on empirical quantities will be even worse. 

The coverages of the statistics based on Monte Carlo simulation are almost identical to those of the statistics based on analytically computed quantities in the spatial Gaussian random field setting. In the multivariate probit model it is not possible to compute the sensitivity and variability matrices analytically, but the statistics based on simulation provide coverages closer to the nominal values than the empirically estimated ones. A further simulation study shows that the computational burden deriving from the simulation of the matrices $H(\theta)$ and $J(\theta)$ can be reduced since $M=500$ repetitions, or even $M=250$, may be enough. Moreover, such moderate number of repetitions can be done in parallel, thus substantially reducing computational time.
In general, it seems that simulation based quantities are preferable, even when the number of independent repetitions of the data is quite large. Therefore, also considering the computational cost of exact calculation of matrix $J(\theta)$ in complex models, the simulation approach should be the default choice whenever simulation from the full model is feasible.

Of course, simulations may be performed also to estimate directly the sampling distribution of the unadjusted composite likelihood ratio statistic, although this procedure can be computationally substantially more demanding. An experiment performed in the two models considered yielded empirical coverages of 94.2\% and 98.9\% for nominal values 95\% and 99\%, respectively, in the spatial Gaussian random field with $n=5$, and coverages 93.8\% and 98.9\%, respectively, in the multivariate probit case when the parameter of interest is $\rho$ and $n=10$. These results are based on 1,000 data sets because the computational cost of this procedure is much higher than the cost for the computation of the adjustment of the composite likelihood ratio statistic. Indeed, the estimation of the distribution of the composite likelihood ratio statistic requires the computation of global and constrained maximum composite likelihood estimates in a large number of simulated data sets, given that the minimum reasonable number of replications for this bootstrap approach is at least 1,000. In an exemplifying case of the Gaussian random field with $n=5$ and $q=64$, a single computation of the composite likelihood ratio statistic took 0.68 seconds; this value multiplied by 1,000 gives approximately 11 minutes, which is the time necessary to compute the bootstrapped sampling distribution of the composite likelihood ratio statistic based on 1,000 repetitions, while the evaluation of the simulated adjustment of the composite likelihood ratio statistic requires only 1.7 seconds circa.

\section*{Acknowledgements}

The authors thank Luigi Pace and Alessandra Salvan for help in the Gaussian random field example, and two referees for their comments which led to an improvement of the paper.
The first author acknowledges the ``Progetto Giovani Ricercatori'' of the University of Padova for financial support of the research project ``Inferential issues in regression models for dependent categorical and discrete data'' carried out at the Department of Statistical Sciences, University of Padova. The second author acknowledges the financial support of the CARIPARO Foundation Excellence - grant 2011/2012.

\newpage
\appendix

\section{Spatial Gaussian random fields}

\setcounter{table}{0}
\renewcommand\thetable{\Alph{section}.\arabic{table}}

The elements of the matrix $J(\theta)$ computed analytically are the following:
{\small
\begin{eqnarray*}
J_{\mu \mu} &=& n \sum_{k>j} \sum_{m>l}  \frac{w_{jk} w_{lm} } {\sigma^{2} (1+\rho_{jk}) (1+\rho_{lm})} \left( \rho_{jl} + \rho_{jm} + \rho_{kl} +\rho_{km}\right), \\
J_{\sigma^{2} \sigma^{2}} &=& n \sum_{k>j} \sum_{m>l}  \frac{w_{jk} w_{lm} } {(\sigma^{2})^{2}} \left\{-1+  \frac{E(A_{ijk}, A_{ilm})}{4  (\sigma^{2})^{2}(1-\rho_{jk}^{2}) (1-\rho_{lm}^{2})} \right\}, \\
J_{\mu\sigma^{2}}&=&J_{\mu\gamma}=0, \\
J_{\sigma^{2} \gamma}&=& \frac{n}{2 \sigma^{2}} \sum_{k>j} \sum_{m>l}   w_{jk} w_{lm} \frac{\partial \rho_{jk}}{\partial \gamma} \frac{1}{(1-\rho_{jk}^{2})(1-\rho_{lm}^{2})} \\
&& \left\{ 2 \rho_{jk} (1-\rho_{lm}^{2})  - \frac{\rho_{jk} E(A_{ijk}, A_{ilm})}{(\sigma^{2})^{2}(1-\rho_{jk}^{2})}+\rho_{jkll}+\rho_{jkmm}-2 \rho_{lm} \rho_{jklm}\right\}, \\
J_{\gamma\gamma} & = & n \sum_{k>j} \sum_{m>l} w_{jk}w_{lm}\frac{\partial \rho_{jk}}{\partial\gamma}
\frac{\partial \rho_{lm}}{\partial\gamma^\top}\frac{1}{(1-\rho^2_{jk})}\frac{1}{(1-\rho^2_{lm})}\\
&&\{-\rho_{jk}\rho_{lm}+\rho_{jklm}-\frac{\rho_{jk}}{(1-\rho^2_{jk})}\left(
\rho_{jjlm}+\rho_{kklm}-2\rho_{jk}\rho_{jklm}\right)-\frac{\rho_{lm}}{(1-\rho^2_{lm})} \\
&& \left(
\rho_{lljk}+\rho_{mmjk}-2\rho_{lm}\rho_{jklm}\right) +\frac{\rho_{jk}}{(1-\rho^2_{jk})}\frac{\rho_{lm}}{(1-\rho^2_{lm})}\\
&& \left(  
\rho_{jjll}+\rho_{jjmm}+\rho_{kkll}+\rho_{kkmm}+4\rho_{jk}\rho_{lm}\rho_{jklm} \right. \\
&& -2\rho_{lm}\rho_{lmjj}
\left. -2\rho_{lm}\rho_{lmkk}-2\rho_{jk}\rho_{jkll}-2\rho_{jk}\rho_{jkmm}
\right)\},
\end{eqnarray*}
where $w_{jk}=w_{ij, ik}$, $\rho_{jklm}=\rho_{jk} \rho_{lm} +\rho_{jl} \rho_{km}+\rho_{jm} \rho_{kl}$ and $E(A_{ijk}, A_{ilm})= (\sigma^{2})^{2} (\rho_{jjll}+\rho_{jjmm}+\rho_{kkll}+\rho_{kkmm}-2\rho_{lm}\rho_{jjlm}-2\rho_{lm}\rho_{kklm}-2 \rho_{jk}\rho_{jkll}-2 \rho_{jk}\rho_{jkmm}+4 \rho_{jk}\rho_{lm}\rho_{jklm})$.

The elements of the matrix $H(\theta)$ are
\begin{eqnarray*}
H_{\mu\mu} & = & \frac{2n}{\sigma^2} \sum_{k>j} \frac{w_{jk}}{1+\rho_{jk}}, \\
H_{\mu\sigma^2} & = &0, \\
H_{\mu\gamma} & = & 0,\\
H_{\sigma^2\sigma^2} & = & \frac{n}{\sigma^4} \sum_{k>j} w_{jk} , \\
H_{\sigma^2 \gamma} & = & -\frac{n}{\sigma^2} \sum_{k>j} w_{jk}\frac{\partial \rho_{jk}}{\partial\gamma}\frac{\rho_{jk}}{(1-\rho^2_{jk})},\\
H_{\gamma\gamma}&=&n \sum_{k>j} w_{jk} \frac{\partial \rho_{jk} }{\partial \gamma} \frac{\partial \rho_{jk} }{\partial \gamma^{\top}} \frac{1+\rho_{jk}^{2}}{(1-\rho_{jk}^{2})^{2}}.\\
\end{eqnarray*}}

Table \ref{tabs1} reports the coverages of the Wald-type and score-type statistics, besides those of the statistics reported in the paper, for the Gaussian random field example. Table \ref{tabM} shows the performance of the simulation based quantities for increasing values of the Monte Carlo repetitions $M$.
\begin{table}[ht]
\begin{center}
{\footnotesize
\begin{tabular}{cccccccccccccc}
$n$		& $LRT$ &	$W^{A}$	&	$W^{E}$	&	$W^{S}$	&	$S^{A}$	&	$S^{E}$	&	$S^{S}$	&	$LR_{2}^{A}$ & 	$LR_{2}^{E}$	& $LR_{2}^{S}$	& $LR_{I}^{A}$	& $LR_{I}^{E}$	&	$LR_{I}^{S}$\\
\hline
\multicolumn{14}{c}{95.0}\\
\hline
1 & 97.1 & 95.7 & - & 95.1 & 97.1 &  -& 97.3 & 99.7 &  -& 99.8 & 98.4 &  -& 98.4 \\ 
  5 & 95.1 & 87.1 & 63.5 & 87.9 & 95.8 & 93.2 & 96.0 & 97.3 & 84.5 & 97.5 & 96.9 & 87.5 & 96.9 \\ 
  30 & 94.8 & 93.2 & 73.9 & 93.3 & 94.9 & 86.7 & 95.1 & 95.3 & 89.1 & 95.5 & 95.1 & 84.3 & 95.2 \\ 
  	\hline			
\multicolumn{14}{c}{99.0}\\
\hline
  1 & 99.7 & 99.1 &  -& 98.7 & 99.7 & - & 99.7 & 99.9 &  -& 100.0 & 99.9 & - & 99.9 \\ 
  5 & 99.1 & 93.5 & 69.2 & 94.4 & 99.2 & 96.4 & 99.3 & 99.4 & 98.7 & 99.3 & 99.3 & 94.4 & 99.3 \\ 
  30 & 98.9 & 97.8 & 81.1 & 97.8 & 99.0 & 94.1 & 99.0 & 99.0 & 96.0 & 99.1 & 99.1 & 92.3 & 99.1 \\ 
   \hline
\end{tabular}}
\caption{Empirical coverages of the statistics: likelihood ratio test based on the ordinary log likelihood ($LRT$), Wald-type ($W$), score-type ($S$), composite likelihood ratio using second order matching adjustment ($LR_{2}$) and composite likelihood ratio adjustment by Pace {\em et al.} (2011) ($LR_{I}$) for nominal values 95\% and 99\% in a spatial Gaussian random field for parameter of interest $(\lambda, \alpha)$, with $n=1, 5, 30$, using analytical ($^{A}$), empirical ($^{E}$) and Monte Carlo ($^{S}$) versions of $H(\theta)$ and $J(\theta)$.}
\label{tabs1}
\end{center}
\end{table}

\begin{table}[tb]
\begin{center}
\footnotesize{
\begin{tabular}{rrrrrrrrrrrrrrrr}
  \hline
  	&	\multicolumn{3}{c}{$W^{S}$} && \multicolumn{3}{c}{$S^{S}$} && \multicolumn{3}{c}{$LR_{2}^{S}$} && \multicolumn{3}{c}{$LR_{I}^{S}$}\\
	\cline{2-4} \cline{6-8} \cline{10-12} \cline{14-16}
 n & 1 & 5 & 30 && 1 & 5 & 30 && 1 & 5 & 30 && 1 & 5 & 30 \\ 
  \hline
  M	& \multicolumn{15}{c}{95.0}\\
  \hline
100 & 94.4 & 87.3 & 92.9 && 96.8 & 95.4 & 94.8 && 99.6 & 97.4 & 95.5 && 98.0 & 96.4 & 95.0 \\ 
  250 & 94.9 & 87.8 & 93.2 && 97.3 & 95.9 & 94.9 && 99.7 & 97.6 & 95.4 && 98.3 & 96.7 & 95.0 \\ 
  500 & 95.1 & 87.8 & 93.3 && 97.3 & 96.0 & 94.9 && 99.7 & 97.6 & 95.5 && 98.3 & 96.9 & 95.1 \\ 
    \hline
  	& \multicolumn{15}{c}{99.0}\\
  \hline
  100 & 98.4 & 94.2 & 97.7 && 99.6 & 99.1 & 98.8 && 100.0 & 99.3 & 99.1 && 99.8 & 99.2 & 98.9 \\ 
  250 & 98.5 & 94.3 & 97.7 && 99.7 & 99.2 & 98.9 && 99.9 & 99.5 & 99.1 && 99.8 & 99.3 & 98.9 \\ 
  500 & 98.7 & 94.4 & 97.8 && 99.7 & 99.3 & 99.0 && 100.0 & 99.4 & 99.2 && 99.8 & 99.3 & 99.1 \\ 
   \hline
\end{tabular}}
\caption{Comparison of coverages of the statistics: Wald-type ($W$), score-type ($S$), composite likelihood ratio using second order matching adjustment ($LR_{2}$) and composite likelihood ratio adjustment by Pace {\em et al.} (2011) ($LR_{I}$) based on Monte Carlo simulation as $M$ increases in a Gaussian random field.}
\label{tabM}
\end{center}
\end{table}

\newpage
\section{Multivariate probit model}
Tables \ref{tabs4}-\ref{tab:covmp2A} report the coverages of the different statistics in the multivariate probit example.
\setcounter{table}{0}

\begin{table}[htdp]
\begin{center}
{\small
\begin{tabular}{c|cccccccc}
$n$	&	$W^{E}$	&	$W^{S}$	&	$S^{E}$	&	$S^{S}$	&	$LR_{2}^{E}$	& $LR_{2}^{S}$	&  $LR_{I}^{E}$	&	$LR_{I}^{S}$\\
\hline
\multicolumn{9}{c}{95.0}\\
\hline
10	&	68.1	&	91.8	&	91.5	&	94.9	&	99.0	&	94.7	&	98.0	&	94.9\\	
30	&	82.7	&	93.5	&	91.7	&	94.7	&	99.0	&	94.6	&	97.7	&	94.7\\
100	&	89.6	&	94.6	&	94.3	&	95.3	&	95.8	&	95.0	&	96.1	&	95.1\\
\hline
\multicolumn{9}{c}{99.0}\\
\hline
10	&	75.7	&	96.1	&	100.0&	98.7	&	99.2	&	98.8	&	99.5	&	98.9\\
30	&	88.6	&	97.8	&	97.2	&	98.9	&	100.0&	98.7	&	99.6	&	99.0\\
100	&	94.8	&	98.6	&	98.6	&	99.0	&	99.4	&	98.8	&	99.3	&	99.1\\	
\hline
\end{tabular}}
\end{center}
\caption{Empirical coverages of the statistics: Wald-type ($W$), score-type ($S$), composite likelihood ratio using second order matching adjustment ($LR_{2}$) and composite likelihood ratio adjustment by Pace {\em et al.} (2011) ($LR_{I}$) for nominal values 95\% and 99\% for the parameter of interest $(\beta_{0}, \beta_{1}, \rho)$ in a multivariate probit model with $q=30$ and $n=10, 30, 100$, using empirical ($^{E}$) and Monte Carlo ($^{S}$) versions of $H(\theta)$ and $J(\theta)$.}
\label{tabs4}
\end{table}%

\begin{table}[d]
\begin{center}
{\small
\begin{tabular}{c|cccccccc}
$n$	&	$W^{E}$	&	$W^{S}$	&	$S^{E}$	&	$S^{S}$	&	$LR_{2}^{E}$	& $LR_{2}^{S}$	&  $LR_{I}^{E}$	&	$LR_{I}^{S}$\\
\hline
\multicolumn{9}{c}{95.0}\\
\hline
10	&	86.0	&	92.8	&	93.6	&	95.1	&	97.8	&	95.6	&	96.1	&	95.1\\	
30	&	90.6	&	93.7	&	93.2	&	94.7	&	97.2	&	95.0	&	96.0	&	94.7\\
100	&	92.8	&	94.6	&	94.7	&	95.1	&	95.3	&	94.9	&	95.4	&	95.1\\
\hline
\multicolumn{9}{c}{99.0}\\
\hline
10	&	90.9	&	96.5	&	99.8	&	99.1	&	99.5	&	99.1	&	99.0	&	99.1\\
30	&	94.8	&	97.8	&	98.4	&	98.9	&	99.4	&	98.9	&	99.1	&	98.9\\
100	&	96.8	&	98.5	&	98.8	&	98.9	&	99.2	&	98.9	&	98.9	&	98.9\\	
\hline
\end{tabular}}
\end{center}
\caption{Empirical coverages of the statistics: Wald-type ($W$), score-type ($S$), composite likelihood ratio using second order matching adjustment ($LR_{2}$) and composite likelihood ratio adjustment by Pace {\em et al.} (2011) ($LR_{I}$) for nominal values 95\% and 99\% for the parameter of interest $(\beta_{1}, \rho)$ in a multivariate probit model with $q=30$ and $n=10, 30, 100$, using empirical ($^{E}$) and Monte Carlo ($^{S}$) versions of $H(\theta)$ and $J(\theta)$.}
\label{tab:covmp1A}
\end{table}%

\begin{table}[htdp]
\begin{center}
{\small
\begin{tabular}{c|cccccccc}
$n$	&	$W^{E}$	&	$W^{S}$	&	$S^{E}$	&	$S^{S}$	&	$LR_{2}^{E}$	& $LR_{2}^{S}$	&  $LR_{I}^{E}$	&	$LR_{I}^{S}$\\
\hline
\multicolumn{9}{c}{95.0}\\
\hline
10	&	89.9	&	93.7	&	89.9	&	94.7	&	97.0	&	95.0	&	97.0	&	95.0	\\
30	&	92.8	&	95.2	&	93.2	&	95.1	&	96.4	&	95.1	&	96.4	&	95.1\\
100	&	94.3	&	94.7	&	94.5	&	95.0	&	95.1	&	95.1	&	95.1	&	95.1\\
\hline
\multicolumn{9}{c}{99.0}\\
\hline
10	&	93.3	&	96.4	&	98.1	&	99.2	&	99.0	&	99.2	&	99.0	&	99.2\\
30	&	96.1	&	98.0	&	97.9	&	99.0	&	99.2	&	99.1	&	99.2	&	99.1\\
100	&	97.4	&	98.6	&	98.8	&	99.0	&	99.0	&	99.0	&	99.0	&	99.0\\	
\hline
\end{tabular}}
\end{center}
\caption{Empirical coverages of the statistics: Wald-type ($W$), score-type ($S$), composite likelihood ratio using second order matching adjustment ($LR_{2}$) and composite likelihood ratio adjustment by Pace {\em et al.} (2011) ($LR_{I}$) for nominal values 95\% and 99\% for the parameter of interest $\rho$ in a multivariate probit model with $q=30$ and $n=10, 30, 100$, using empirical ($^{E}$) and Monte Carlo ($^{S}$) versions of $H(\theta)$ and $J(\theta)$.}
\label{tab:covmp2A}
\end{table}%

\end{document}